# SIR MODEL FOR COVID-19 CALIBRATED WITH EXISTING DATA AND PROJECTED FOR COLOMBIA

Carlos Armando De Castro[1]
engineering@cadecastro.com

## 1. INTRODUCTION

Since January of 2020 there's an epidemic of COVID-19, a disease caused by the novel coronavirus Sars-CoV-2 that started in a market at Wuhan, China and spread to the whole world, at which point it was declared a pandemic by the World Health Organization (WHO)[2] and led to the closing of borders and cancellation of flights like never seen before.

In this paper we develop a SIR epidemiological model with parameters calculated according to existing data [1] at the time of writing (**24/03/2020**); the data is from Italy, South Korea and Colombia, the model is then used to project the evolution of the COVID-19 epidemic in Colombia for different scenarios using the data of population for the country and known initial conditions at the start of the simulation [2].

## 2. MATHEMATICAL MODEL

We use the *SIR* (Susceptible-Infected-Removed) model which separates a population into three groups at any given time: susceptible individuals to the disease (total of people who hasn't been infected) noted by $S$, infected individuals (people with the infection active at the moment) noted by $I$, and removed individuals (recovered or dead) noted by $R$; the non-linear system of first-order ordinary differential equations is [3]:

$$dS/dt = -\beta SI$$

$$dI/dt = \beta SI - \gamma I \quad (1)$$

$$dR/dt = \gamma I$$

A big assumption of this model is that people recovered from COVID-19 will not be re-infected (which hopefully seems to be the case). The initial conditions are $S(0)=S_0$, $I(0)=I_0$ y $R(0)=R_0$; $\beta$ and $\gamma$ are positive constants found by adjustment to existing data. To solve the system, it is only needed to solve the first two equations since $S$ and $I$ do not depend on $R$. The equilibrium (no new infected individuals) is reached when:

$$S_e = \frac{\gamma}{\beta} \quad (2)$$

[1] *Mechanical engineer*. **cadecastro.com**
[2] https://www.who.int/es/emergencies/diseases/novel-coronavirus-2019

## 3. NUMERICAL MODEL AND CALIBRATION

The system (1) is discretized by finite-difference in a Euler implicit method [4] to guarantee stability of the numerical solution, with a unit time step (Δt = 1 day) we get:

$$S_n - S_{n-1} = -\beta S_n I_n$$

$$I_n - I_{n-1} = \beta S_n I_n - \gamma I_n \quad (3)$$

Where the sub-index indicates the time step (in this case, the day) of the simulated variable. Algebraically developing the system (3) we get the values for each variable at day *n*:

$$\boxed{I_n = \frac{\sqrt{[1+\gamma-\beta(S_{n-1}+I_{n-1})]^2 + 4\beta(1+\gamma)I_{n-1}} - [1+\gamma-\beta(S_{n-1}+I_{n-1})]}{2\beta(1+\gamma)}} \quad (4)$$

$$\boxed{S_n = \frac{S_{n-1}}{1+\beta I_n}} \quad (5)$$

The numerical solution was implemented in an Excel spreadsheet. The calibration of the SIR model parameters is done adjusting with the existing data, for day *n* we get solving from the equations:

$$\beta_n = -\frac{(dS/dt)_n}{S_n I_n} \quad (6)$$

$$\gamma_n = \frac{\beta_n S_n I_n - (dI/dt)_n}{I_n} \quad (7)$$

The derivatives are calculated with first-order finite-difference from each data set. Then we take the average value for *N* days of a selected interval for the adjustment:

$$\beta = \frac{\sum_{n=1}^{N} \beta_n}{N} \quad (8)$$

$$\gamma = \frac{\sum_{n=1}^{N} \gamma_n}{N} \quad (9)$$

The value β must be normalized multiplying by the total population to be able to compare.

## 4. PROJECTION FOR COLOMBIA WITH PARAMETERS FROM ITALY

Italy is the hardest hit country in Europe at the time of writing this paper, the epidemic grew initially without control so its parameters are a simulation scenario of interest, the data of epidemic [1] and population [2] for Italy was analyzed and the model adjusted with data between 29/02/2020 and 20/03/2020, results from equations (6) to (9) are the following:

| Population = | 60461826 |
|---|---|
| **β** | **γ** |
| **3.245E-09** | **3.163E-02** |

**Table 4.1**. SIR model parameters calibrated with data from Italy.

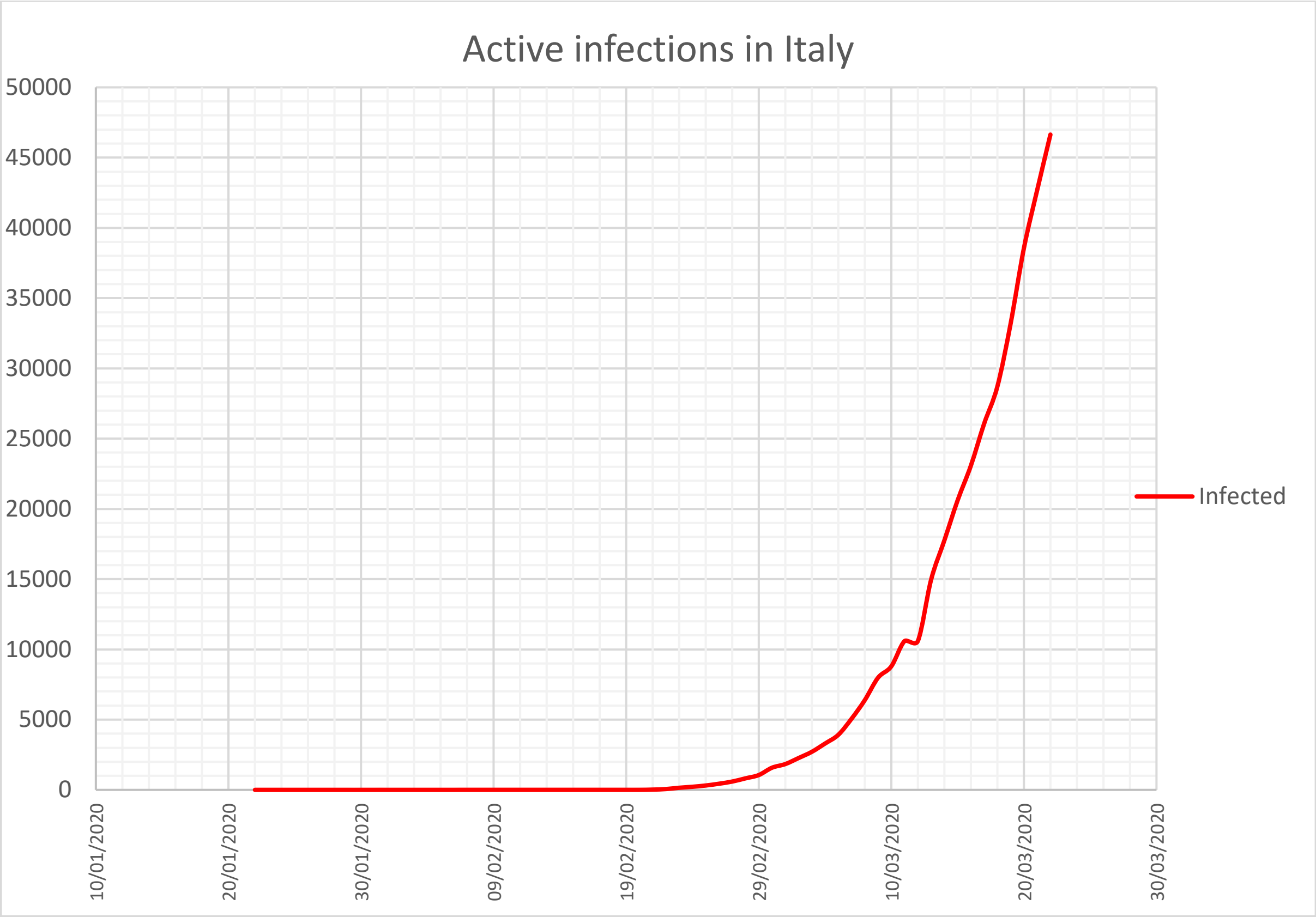

**Figure 4.1**. Active infected people in Italy.

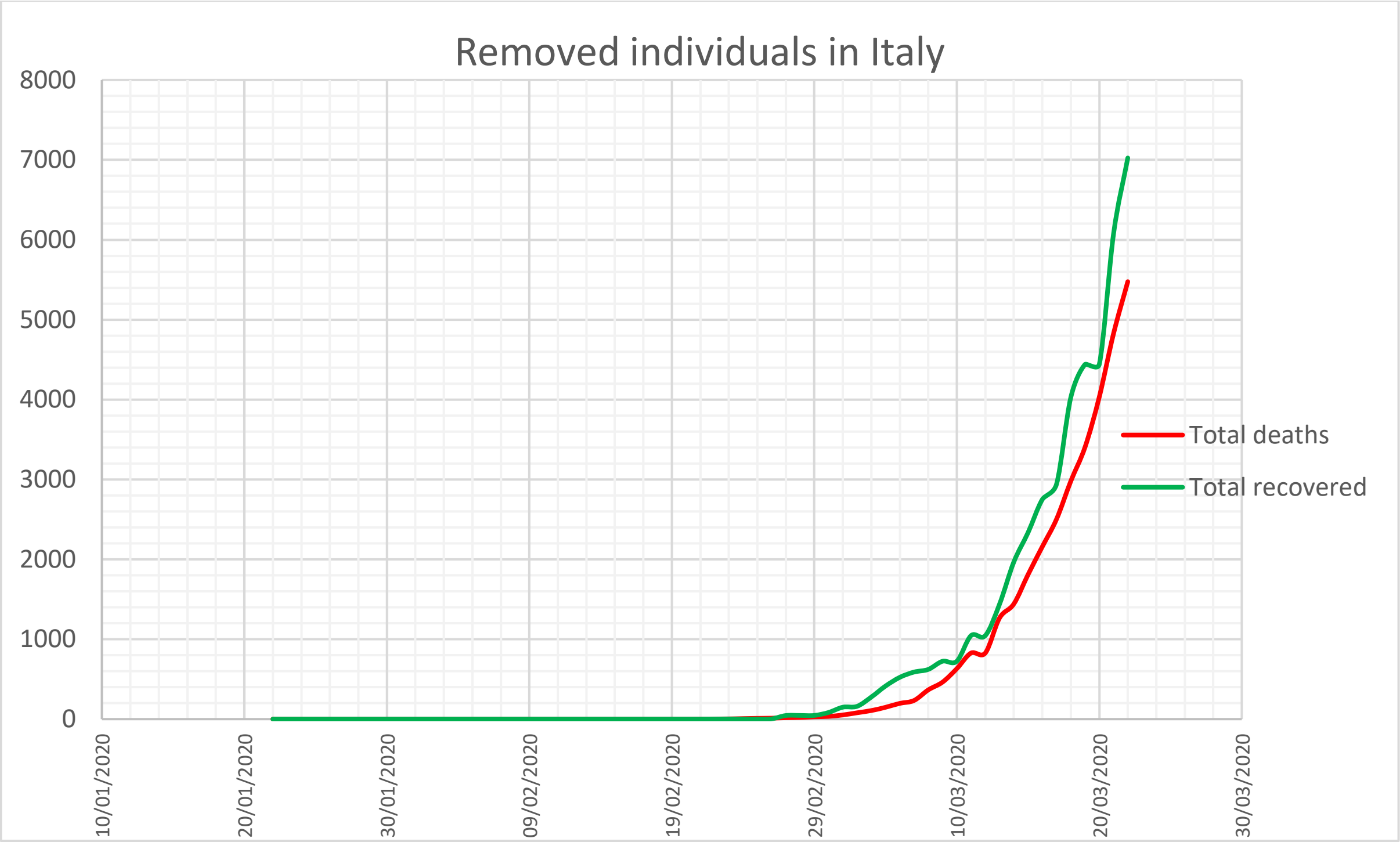

**Figure 4.2.** Total recovered and deaths in Italy.

With these SIR parameters and the population of Colombia [2] we simulate the COVID-19 evolution in the country for a year with $I_0$=369 (by 24/03/2020) with 9 removed individuals (6 recovered and 3 deaths). The results are shown in Figure 4.3:

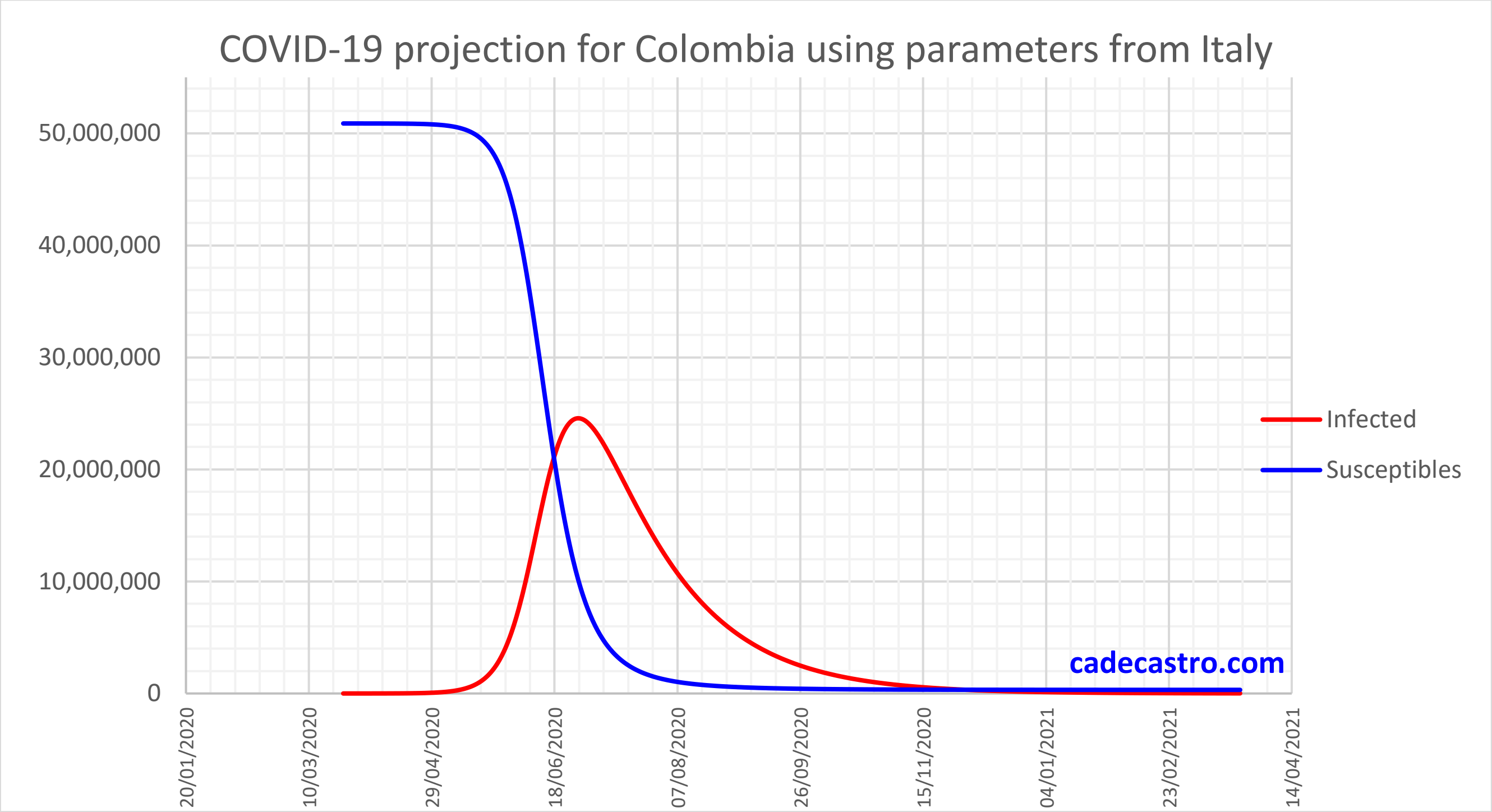

**Figura 4.3.** Simulated infected individuals for Colombia using parameters from the infection in Italy.

It is observed that the peak of infections would be by 29/06/2020 with a total of 25 million infected individuals at the same time. The infection would get low values by December of 2020.

## 5. PROJECTION FOR COLOMBIA WITH PARAMETERS FROM SOUTH KOREA

South Korea has been an example by their discipline testing COVID-19 cases and their control of the epidemic, for uncontrolled growth data was taken between 20/02/2020 and 29/02/2020, for controlled growth data was taken between 10/03/2020 and 20/03/2020.

| Population | 51269185 |
|---|---|
| Uncontrolled | |
| β | γ |
| 7.202E-09 | 8.684E-03 |
| Controlled | |
| β | γ |
| 2.887E-10 | 1.886E-02 |

**Table 5.1**. SIR model parameters calibrated with data from South Korea.

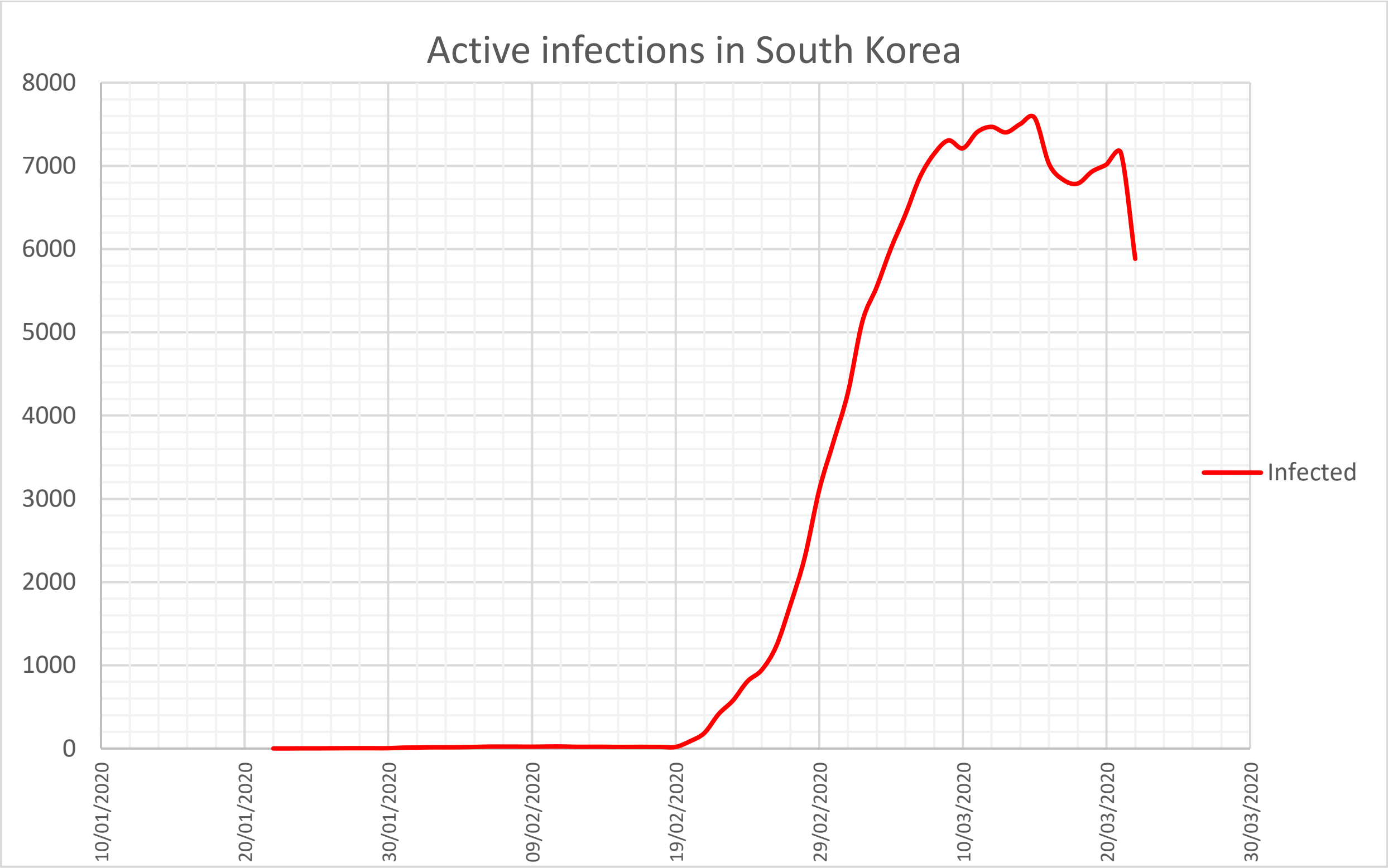

**Figure 5.1**. Active infected people in South Korea.

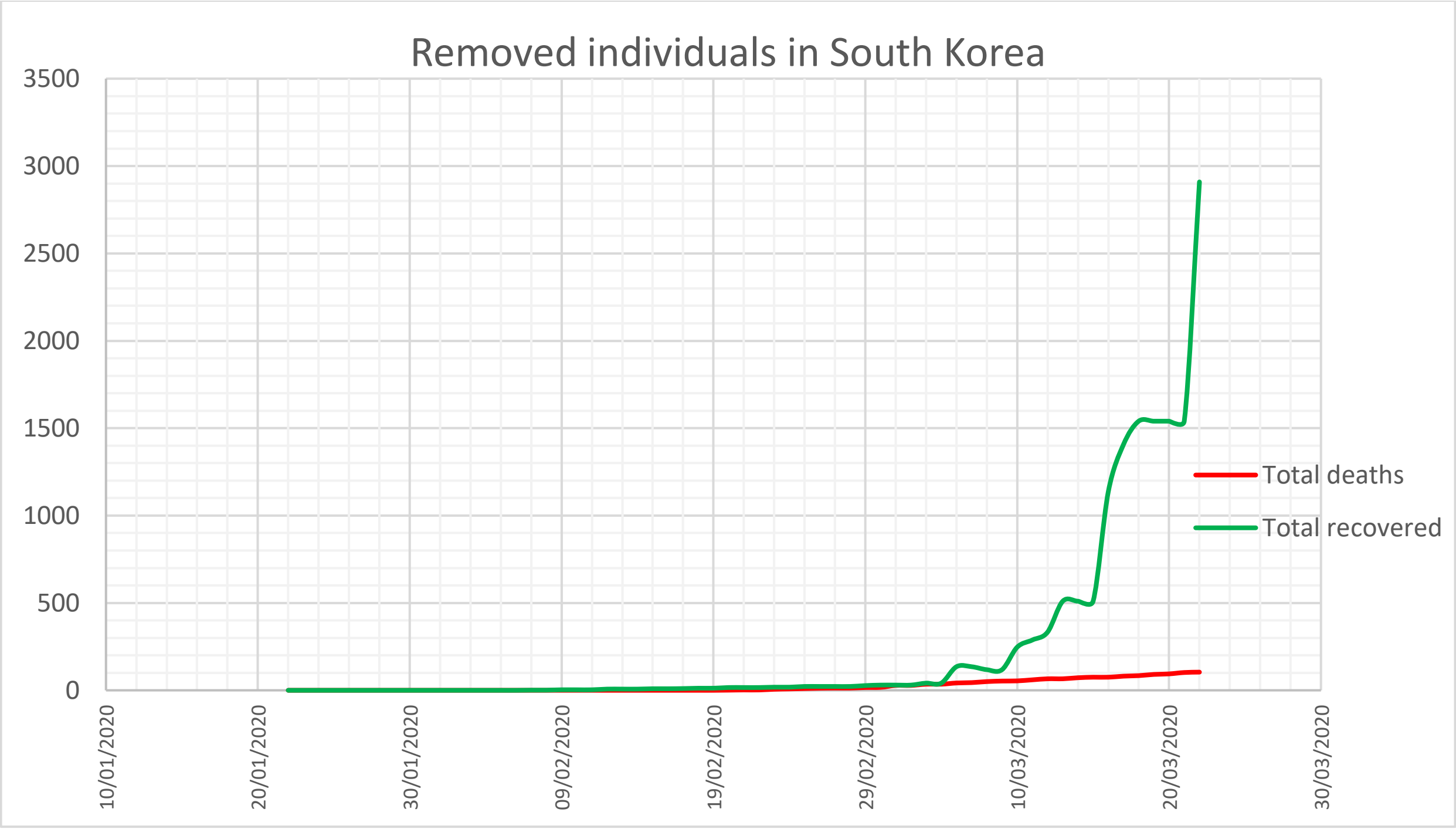

**Figure 5.2.** Total recovered and deaths in South Korea.

With these SIR parameters and the same population and initial conditions from Section 4 we simulate for Colombia:

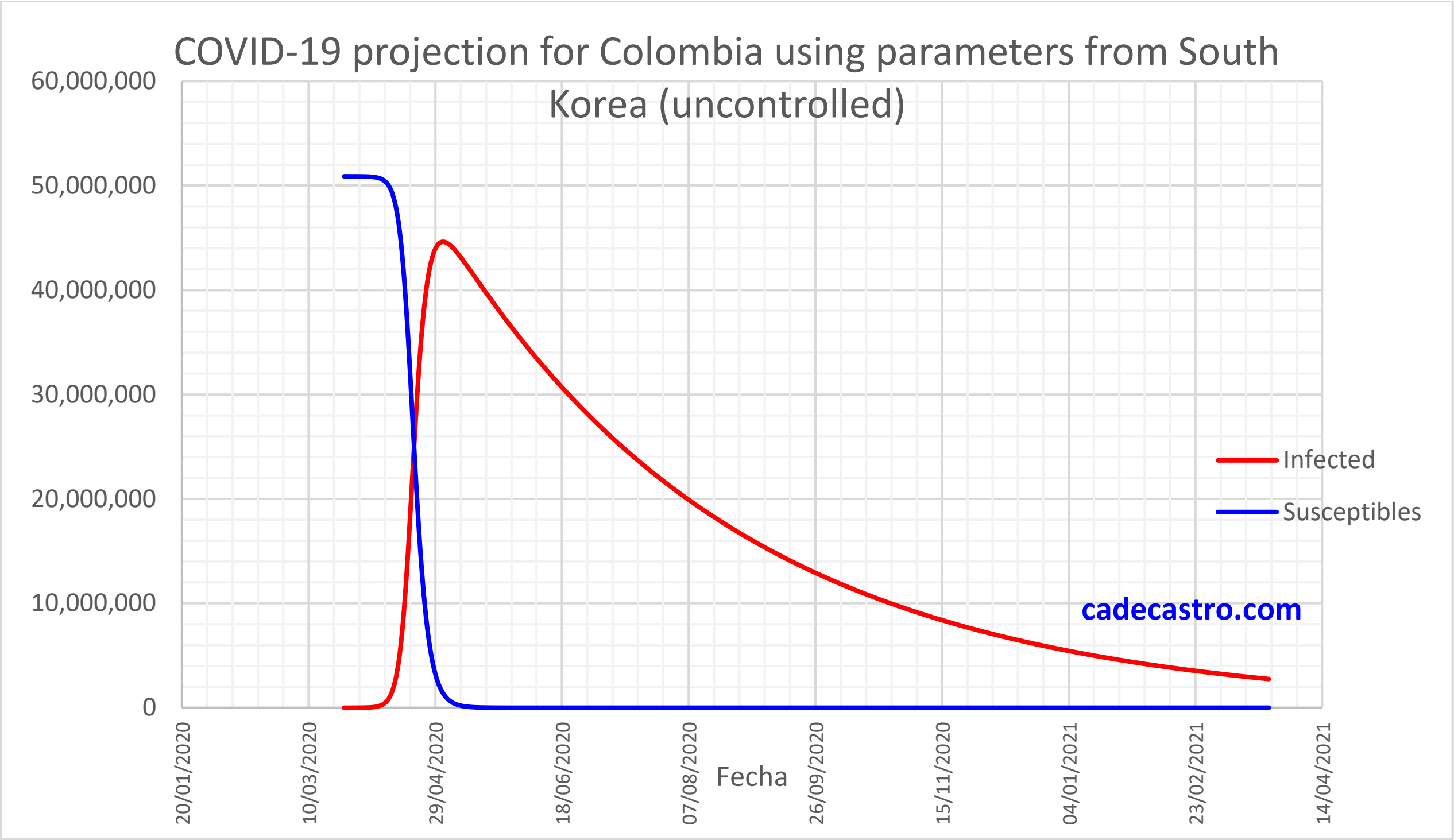

**Figure 5.3.** Simulated infected individuals for Colombia using parameters from the uncontrolled infection in South Korea.

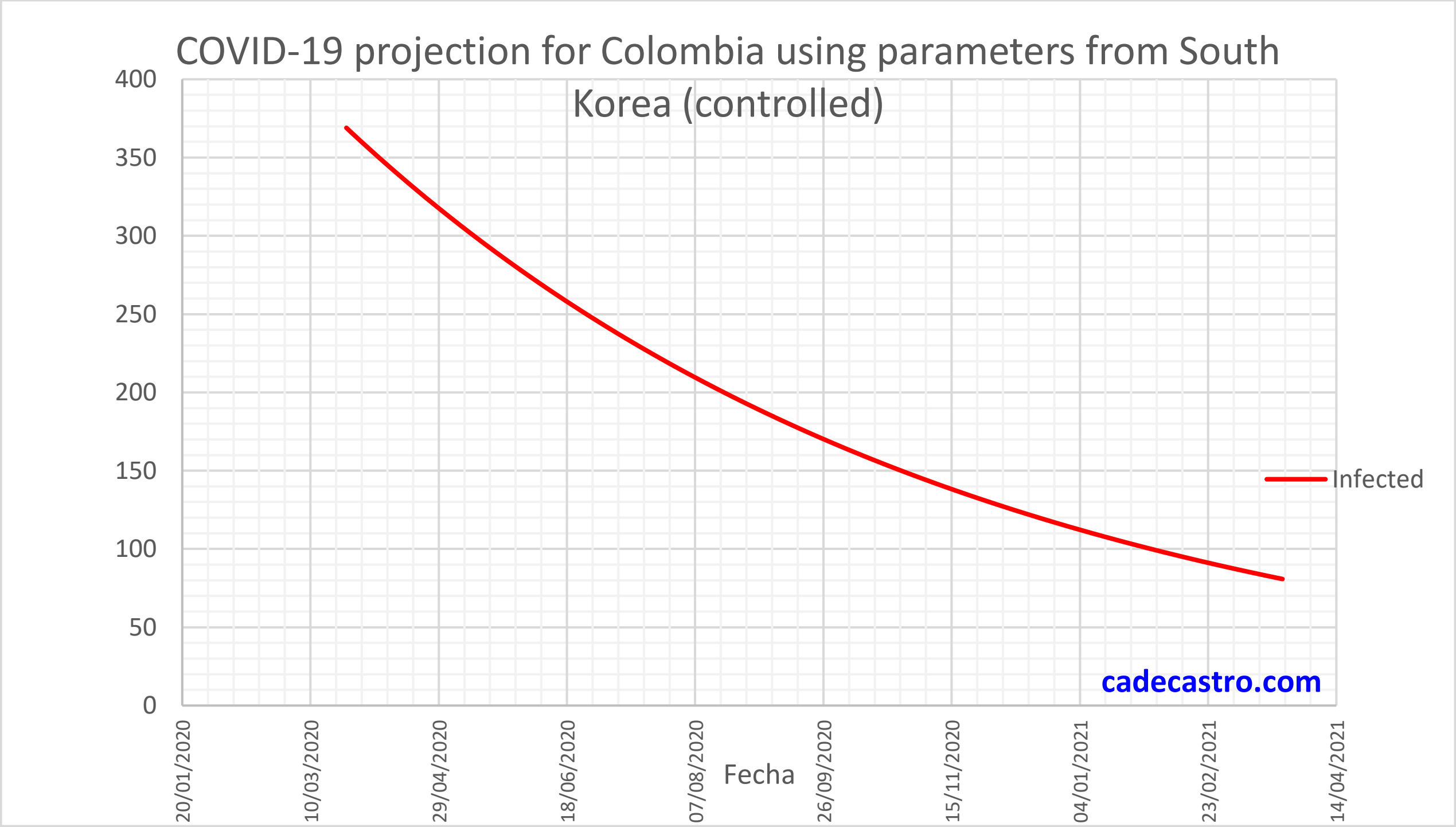

**Figure 5.4.** Simulated infected individuals for Colombia using parameters from the controlled infection in South Korea.

There's a substantial difference between results, the uncontrolled epidemic would reach a peak of 44.6 million infected people by 02/05/2020 (creating an unbearable burden on the Health System and possibly millions of deaths) while the totally controlled situation gives us a declining number of active infections, that would be the most ideal case.

## 6. PROJECTION FOR COLOMBIA WITH OWN PRE-LOCKDOWN DATA

The model is calibrated with existing data from Colombia [1] before a national lockdown that takes effect the day of the writing of this paper, although this data is very limited compared to Italy and South Korea since the infection arrived later on the country.

| β | γ |
|---|---|
| 5.295E-09 | 3.332E-03 |

**Table 6.1**. SIR model parameters calibrated with data from Colombia.

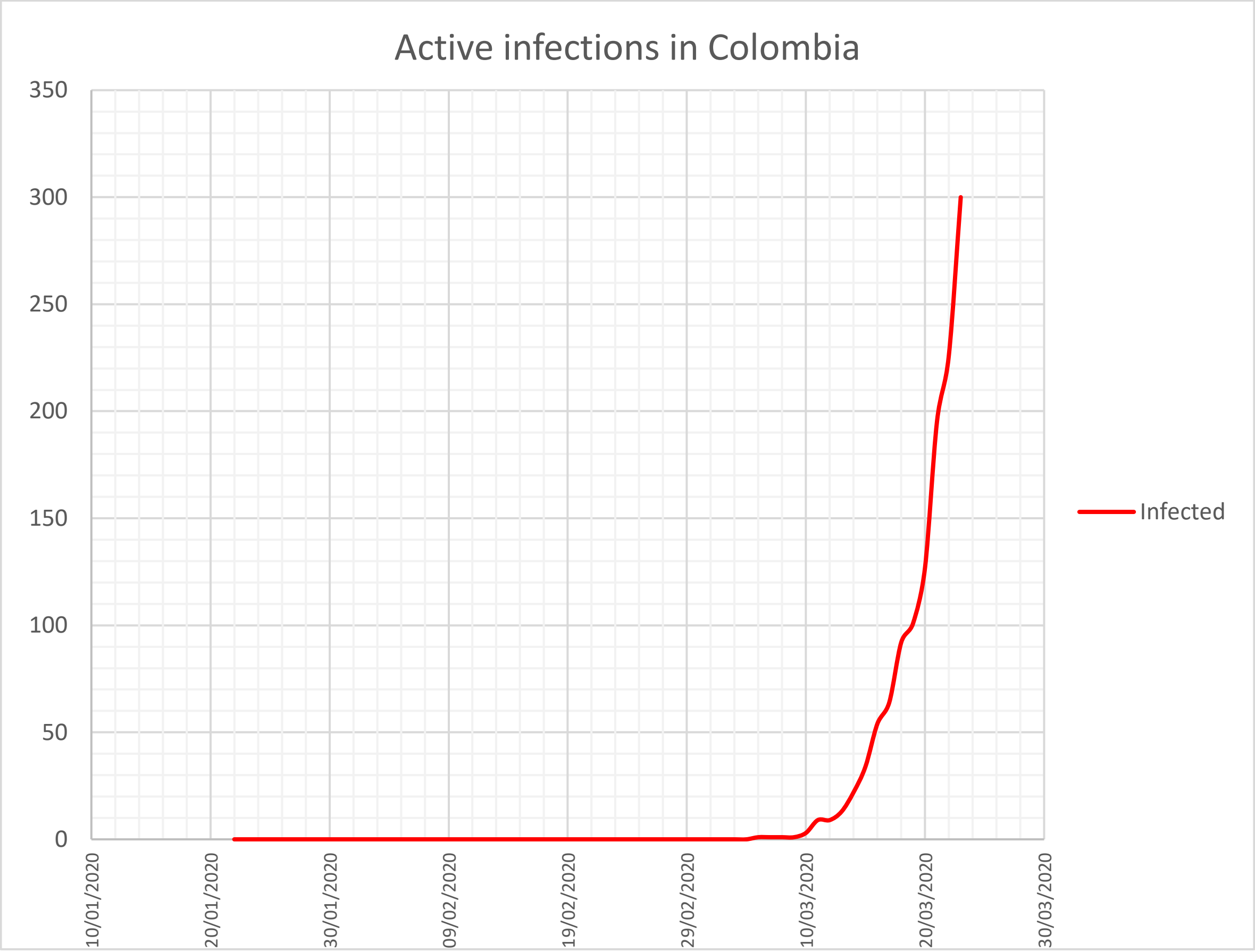

**Figure 6.1**. Active infected people in Colombia.

With these SIR parameters and the same population and initial conditions from Sections 4 and 5 we simulate for the next year in Colombia (Fig. 6.3).

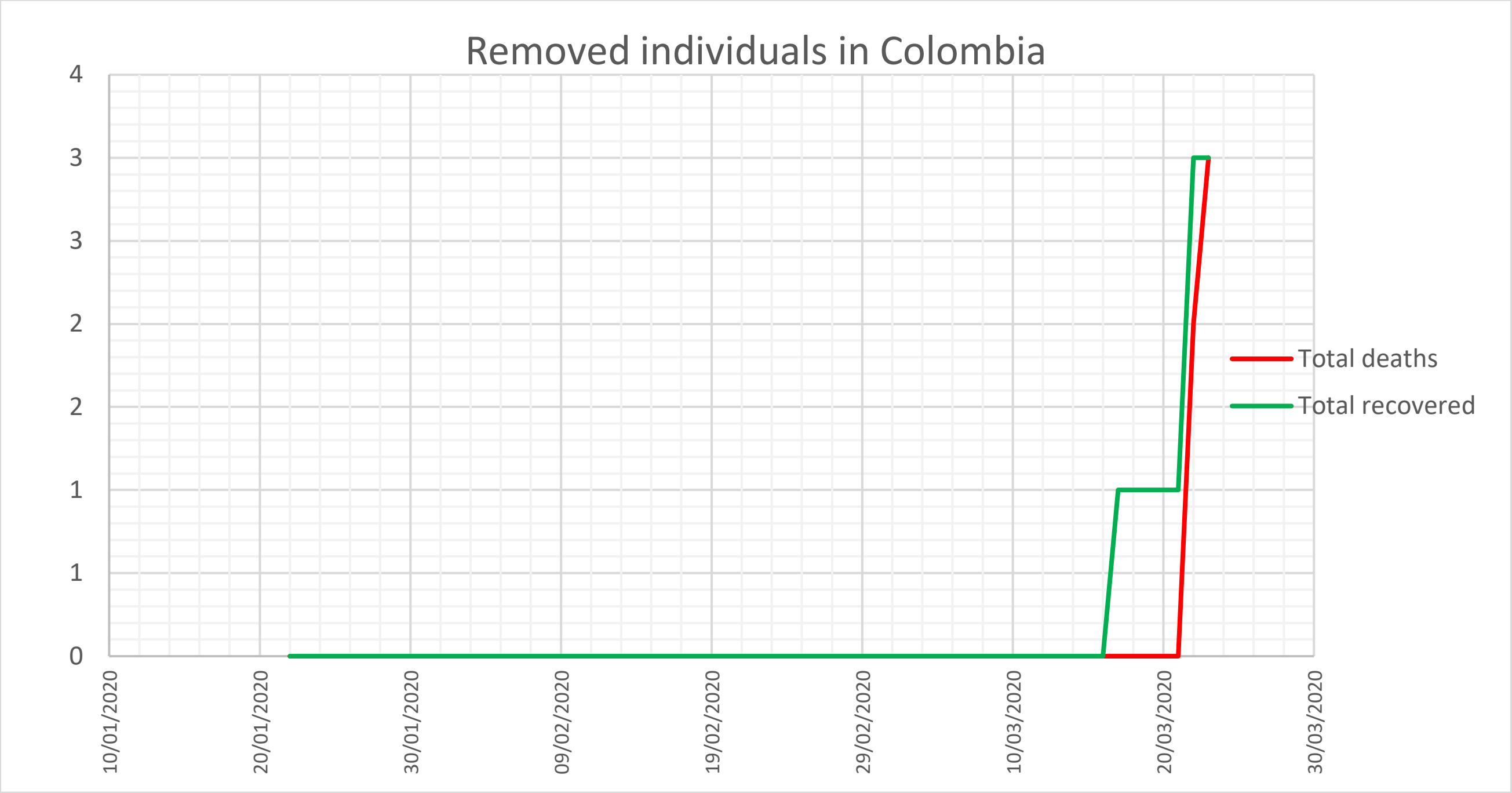

**Figure 6.2.** Total recovered and deaths in Colombia.

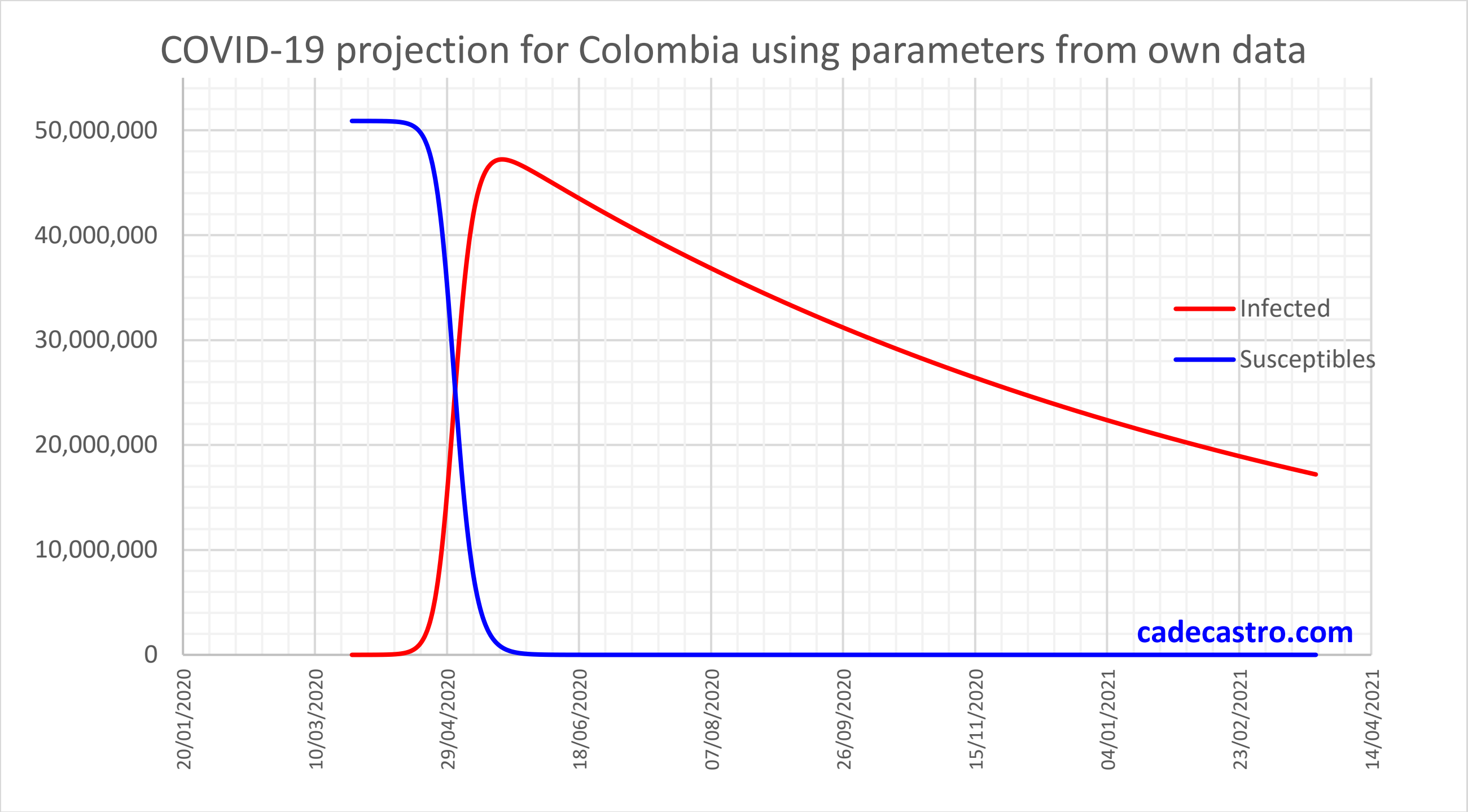

**Figure 6.3.** Simulated infected individuals for Colombia using parameters from the beginning of the infection in the country.

According to the simulation if the infection is left as of right now there would be a huge peak of 47.2 million infected people by 21/05/2020, with millions of infected individuals taking a toll on the Health System for more than a year, a disastrous outcome.

## 7. CONCLUSIONS

We developed a numerical model for simulating the infection of COVID-19 using the data obtained from open sources and solving the SIR model; this numerical model is useful for any region adjusting the corresponding parameters.

The gravest conclusion is that the COVID-19 spreads very easily infecting scores of people in a short time with huge peaks of active infections, thus representing a very extreme threat to Health Systems in all the world (Italy and Spain being examples of this as of the time of this writing).

It seems that the lockdown approach to combat the spread of COVID-19 is adequate from the data from South Korea.

## REFERENCES:

[1] ***COVID-19 Pandemic World Data – open access***. Recovered from https://github.com/CSSEGISandData/COVID-19 on 24/03/2020.

[2] ***World Population by Country***. Recovered from https://www.worldometers.info/world-population/population-by-country/ on 24/03/2020.

[3] Hirsch, Smale & Devaney. ***Differential Equations, Dynamical Systems and an Introduction to Chaos***. Elsevier.

[4] Richard L. Burden, J. Douglas Faires. ***Numerical Analysis***. 5th edition. PWS Publishing Company.